\documentclass[a4paper,UKenglish]{lipics}
\usepackage{zed-csp}
\usepackage[framemethod=default]{mdframed}

\usepackage{lipsum}
\title{Teaching Logic to Information Systems Students: Challenges and Opportunities}
\titlerunning{Teaching Logic and Formal Specification to Information Systems Practitioners} %optional, in case that the title is too long; the running title should fit into the top page column

\author[1]{Anna Zamansky}
\author[2]{Eitan Farchi}

\affil[1]{Information Systems Department, University of Haifa\\
  Haifa, Israel\\
  \texttt{annazam@is.haifa.ac.il}}
  \affil[2]{IBM Research\\
  Haifa, Israel\\
  \texttt{farchi@ibm.il}}
\authorrunning{A. Zamansky and E. Farchi} %mandatory. First: Use abbreviated first/middle names. Second (only in severe cases): Use first author plus 'et. al.'

\Copyright{Anna Zamansky and Eitan Farchi}%mandatory, please use full first names. LIPIcs license is "CC-BY";  http://creativecommons.org/licenses/by/3.0/

% mandatory: Please provide 1-5 keywords
\serieslogo{logo_ttl}%please provide filename (without suffix)
\volumeinfo%(easychair interface)
  {M. Antonia {Huertas}, Jo\~ao {Marcos}, Mar\'ia {Manzano}, Sophie {Pinchinat}, \\
  Fran\c{c}ois {Schwarzentruber}}% editors
  {5}% number of editors: 1, 2, ....
  {4th International Conference on Tools for Teaching Logic}% event
  {1}% volume
  {1}% issue
  {273}% starting page number
\EventShortName{TTL2015}
\begin{document}

\maketitle

\begin{abstract}
In contrast to Computer Science, where the fundamental role of Logic is widely recognized, it plays a practically non-existent role in Information Systems curricula. In this paper we argue that instead of Logic's exclusion from the IS curriculum, a significant adaptation of the contents, as well as teaching methodologies, is required for an alignment with the needs of IS practitioners. We present our vision for such adaptation and report on concrete steps towards its implementation in the design and teaching of a course for graduate IS students at the University of Haifa. We discuss the course plan and present some data on the students' feedback on the course.

\end{abstract}

\section{Introduction}

The fundamental role of Logic in the Computer Science curriculum is widely recognized. The ACM CS undergraduate curriculum guidelines (\cite{sahami2011setting}) explicitly state logic as a mathematical requirement which is directly relevant for the large majority of all CS undergraduates (together with elements of set theory and discrete probability). This recommendation is implemented in most of the standard CS undergraduate study programs by including in the curriculum a course in discrete structures which includes a significant amount of formal logic. 

The (academic) field of Information Systems (IS) encompasses two broad areas: (i) acquisition, deployment, and management of information technology resources and services, and (ii) development and evolution of infrastructure and systems for use in organization processes. Thus, as opposed to CS,  IS's primary focus is on an organization’s mission and objectives and the application of information technology to further these goals. Yet both IS and CS require a common subset of technical knowledge, reflected also in the intersection of the respective study programs' curricula.

Logic, however, does not appear to be in this intersection -- almost none of the IS undergraduate study programs include such course in their curriculum. The ACM IS curriculum guidelines (\cite{topi20102010}), which mention statistics and probability as required core IS topics and discrete mathematics as an optional one, but do not refer to logic as relevant: {\em ``Even though IS professionals do not need the same level of
mathematical depth as many other computing professionals, there are, however, some
core elements that are very important for IS professionals. To support in-depth analysis of data, IS
professionals should have a strong background in statistics and probability. For those
who are interested in building a strong skill
set in algorithmic thinking, discrete mathematics
is important."}

We believe the current state of affairs is suboptimal for several reasons. First of all, all the reasons for including logic in the CS curricula still hold in the IS domain: it is widely acknowledged that studying logic directly contributes to software development skills, confirmed recently also by empirical studies (\cite{page2003software}). Secondly, the lack of experience with formal notation forms a major cognitive barrier to the adoption of formal methods in the industry (\cite{zimmerman2002investigating}). This is further reinforced by the fact that 
because many IS study programs tend to be marketed as programs "excluding the hard math"\footnote{Below are some exemplary quotes from webpages of academic institutions providing both CS and IS degrees on the comparison between the two: \begin{itemize}\item ``As a rule, computer science requires more mathematics and analytical skill than information systems. Also, our experience has shown that it is easier to move from being a CS major to an IS major than the other way around.  Therefore, if you feel reasonably comfortable with the math requirements, then you should start out as a computer science major." (Saint Michael's College) 
\item ``A major in CS will know a considerable amount of mathematics which will help in technological applications...An IS major needs to be aware of what information technology can contribute to an organization and how to bring that solution to fruition." (University of Missouri-St. Louis).
\end{itemize}}, the students come to see the lack of mathematical courses as a benefit, and express disappointment\footnote{Quoting one of our graduate students who was assigned to read a research paper on formal methods: ``When I see formal definitions, I just want to cry." Notably, she is one of the best students in her class.} when any formal notations are integrated in the IS core courses -- thus creating a vicious circle.  

%adoption of formal methods by organizations has a great potential of helping meet IS objectives. However, one of the main factors hindering such adoption is the lack of experience and competence in applying formal notation by practitioners (\cite{zimmerman2002investigating}). The lack of logic in the IS curriculum contributes to the reinforcement of this factor. Moreover, as 

Although we tend to agree that a typical IS major may need a less extensive mathematical background than a CS major, we believe that rather than excluding 
logic from the IS curriculum, a significant adaptation of the contents that are taught to align with IS objectives. Recently more voices are calling for a reconsideration of the traditional logic curriculum and its adaptation to the needs of future pracitioners. Several proposals on {\em what} to teach and {\em how} to teach it have been made in the context of CS (\cite{makowsky2008hilbert,tavolato2012integrating,DBLP:conf/amast/Wing95}).

%More importantly, we believe that logic should not be seen as a general mathematical background for the core topics of IS -- it is in the heart of almost each and every one of them, and as such should be integrated in the IS curriculum. Yet a change of emphasis is required to adapt the content of logic courses to the IS domain. 
In this paper we address these questions in the context of the IS curriculum. We report on our experience in 
designing and teaching the course "Logic and Formal Specification" to graduate students at the Information Systems (IS) department at the University of Haifa, which is one of the few to include a {\em mandatory} course on logic and formal methods in its graduate study program. We discuss our view of {\em what} should be included in the ``IS logic toolbox" in order for the students to be able to carry out activities for checking (by proof, analysis or testing) that a software system meets specifications and fulfills its intended purpose. These tools should {\em at least}
(i) by providing background on induction and propositional and first-order logic, and (ii) by providing the ability to read, write and reason about formal specifications. We share our insights on {\em how} the above can be achieved: (i) excluding complex mathematical intricacies, and (ii) providing simple yet software-related examples. Concerning the latter, we report on an ongoing work to develop a tool for measuring ``simplicity" of Z specifications. While the above insights are yet to be empirically validated, some indications of the benefits of our approach are reflected in our students' feedback, which we briefly discuss below.

\section{The IS Logic Toolbox}

The main practical objective in teaching logic to IS practitioners is to give them the ability to apply formal methods in industry. Application of formal aspects is particularly important for software quality control, i.e., activities for checking (by proof, analysis or testing) that a software system meets specifications and that it fulfills its intended purpose. 

Due to the density of the IS curricula, currently one cannot afford to have one course on pure formal logic and then another on formal methods (this problem is also discussed in \cite{wing2000invited} in the context of CS). Therefore, one must develop some mixture of a introductory formal logic together with introduction to formal methods relevant for the IS domain. In what follows we briefly survey previous reflections on the content of logic and formal methods courses that practitioners {\em really} need and their integration into the curricula, and propose how to adapt the proposed ideas for the context of IS.

\subsection{Previous Proposals}

Recently there has been an ongoing discussion on whether the traditional logic syllabus for CS is relevant for practitioners. Since our main goal in this paper is to extend and adapt this discussion to the context of IS, we start by briefly outlining some relevant proposals (mostly in the context of CS), the ideas of which are close in spirit to the vision we present below.

In his paper ``From Hilbert to a Logic Toolbox" \cite{makowsky2008hilbert}, J. Makowsky questions the suitability of the standard logic syllabus to the needs of CS practitioners. He states: 
``The current syllabus is often justified more by the traditional narrative
than by the practitioner’s needs." He further notes that most classical logic textbooks follow the narrative of the rise and fall of Hilbert's program, emphasizing the following ideas: (i) Logic is needed to resolve the paradoxes of set theory;
(ii) First-order logic (FOL) is The logic due to its Completeness theorem; (iii) The main theorems of FOL are the Completeness and Compactness theorems; (iv) The tautologies of FOL are not recursive; (v) One cannot prove consistency within rich enough systems. This, according to Makowsky, is {\em not} what a CS practitioner needs: ``The proof of the Completeness Theorem is a waste of time at the expense of teaching more the important skills of understanding the manipulation and meaning of formulas." What he needs is:  
(i) understand the meaning and implications of modeling the environment as precise mathematical objects and relations; (ii) understand and be able to distinguish intended properties of this
modeling and side-effects; (iii) be able to discern different level of abstraction, and (iv) understand what it means to prove properties of modeled objects. 
%and relations

%\item master the (non-formalized) language of sets and
%second order
%logic which enables her to speak about the modeled objects;

%\item understand what it means to prove properties of modeled objects
%and relations;

%\item understand the
%inherent limitations
%of what can be
%achieved. 

%\end{itemize}

In her papers \cite{wing2000invited,DBLP:conf/amast/Wing95}, J. Wing stresses the importance of integrating formal methods into the existing CS curriculum by teaching their common conceptual elements, including state machines, invariants, abstraction, composition, induction, specification and verification. She states discrete mathematics and mathematical logic as crucial prerequisites. 

The above proposals on {\em what} to teach are extremely relevant for IS practitioners. On the question of {\em how} to teach, the paper ``Integrating Formal Methods into Computer
Science Curricula at a University of Applied
Science" (\cite{tavolato2012integrating}) of Tavolato and Vogt offers some useful insights. It discusses teaching formal methods at universities of applied sciences, where there are usually limiting factors which are relevant to the IS context as well: (i) students have very limited theoretical background, and (ii) they are strongly focused on the direct applicability of what they are taught. In this context the authors stress the importance of making the practical applicability of the theory understandable to students, and making use of real industry-inspired examples. \\
In what follows, we extend and adapt the above proposals to the context of IS, and provide our vision on aligning the teaching of logic to the needs of IS practitioners.

\subsection{Our Vision: Making Logic Relevant for IS}\label{vision}

Logic is a prerequisite for understanding and successfully using formal methods, which in their turn 
can significantly contribute to software quality control. We agree with \cite{wing2000invited} that the main basic formal conceptual elements that the students need to be familiar with include state machines, abstraction, composition, induction, invariants, specification and verification. While the students encounter the concepts of state machines, abstraction and composition at other IS courses (such as modeling and design), aspects related to working with formal specifications are not covered elsewhere. This leads to the following practical needs of an IS practitioner: (1) read, write and understand formal specifications, (2) be able to formalize informal specifications, (3) analyze specifications and detect sources of incompleteness, inconsistency and complexity, (4) reason about specifications, and (5) check a system against a specification.

Based on the above, adapting and extending the previous proposals to the context of IS, we arrive at the following IS logic toolbox: (a) Basic principles for reasoning about sets; (b) Induction and invariants; (c) Propositional and first-order logic and their axiomatizations;  
(d) Formal specification and verification.

As to {\em how} to teach logic to IS students, i.e., designing concrete teaching methodologies, the following considerations need to be taken into account: \begin{itemize}

\item {\em Examples from software domains are useful}. Although it has been believed for some time that studying logic improves software development skills, this common belief has recently been empirically validated. In a three-year study in the framework of the Beseme project (\cite{page2003software}), empirical data on the achievements of two student populations was collected: 
those who studied discrete
mathematics (including logic) through examples
focused on reasoning about software, and those who studied
the same subject illustrated with more
traditional examples. An analysis of the data revealed significant differences in the programming effectiveness of these two populations in favor of the former. As pointed out by \cite{tavolato2012integrating}, software related examples are also useful for increasing the motivation of students, who can see the applications of the studied material in the domain of their interest.

\item {\em Cognitive difficulties should not be ignored.} Empirical studies show that the use of formal methods poses objective difficulties for practitioners (\cite{carew2005empirical,finney1996mathematical}). They are also hypothesized to be a major hindering factor for the acceptance of formal methods in industry (\cite{zimmerman2002investigating}). Although the cognitive processes of students when studying logic and formal methods are not well understood, they should not be ignored (\cite{tavolato2012integrating}). Numerous studies in the education community addressed the gap between the students' intuition and formal thinking in mathematics (see, e.g., \cite{ejersbo2014bridging}). Implementing similar ideas in the domain of teaching logic and formal methods may help deal with these barriers.

\item {\em Intricate complexities are not always needed}. Exposing the students to full intricate complexities of mathematical logic (such as a full proof of the completeness theorem, or dealing with variables not free for substitution) has the potential to confuse novices struggling to understand new ideas. However, most of the practitioners will not encounter them in industry. This is in line the research agenda of indirect application of formal methods (\cite{hussmann1995indirect}), calling for hiding the intricate complexities behind automatic tools with intuitive user interface. The benefits of hiding logical complexity behind the more intuitive interface of functional programming are also mentioned in \cite{page2003software}.

\end{itemize}

In view of the above considerations, the basic principles in the design of the course described below have been 
(i) use mainly examples from software domains,  (ii) use comprehensible examples, and (iii) introduce the logical concepts at a basic level. The issue of comprehensibility also led to our ongoing work of automatically measuring comprehensibility of Z specifications, which we briefly describe below.

%In what follows we describe how this vision has been implemented in an IS graduate course.  

\section{Teaching Logic at the IS Department of the University of Haifa} \label{Haifa}

In this section we demonstrate how the vision presented in Section \ref{vision} has been implemented in our design of the course ``Logic and Formal Specification". The course has been taught at the IS department at the University of Haifa for several years by both of the authors\footnote{Perhaps it is important to mention here the authors' relevant background. The first author is an associate professor at the Information Systems Department at the University of Haifa with active research interests in applied logic. The second author is the manager of the Software Performance and Quality research group at the IBM Haifa Research Laboratory, and a member of the IBM corporate Board of Software Quality. Both of the authors have several years of experience in teaching logic and formal methods to various audiences of students. }. The course is a mandatory course for graduate students, and its length is one semester, 4 hours per week.

%In what follows we describe the course structure and share our insights on teaching the course to the described audience. From the students' angle, we discuss some data on their feedback, obtained in a preliminary study conducted by the second author in the past two years.    

\subsection{Course Description}
Below we provide a short description of the course's main topics, which are divided into two main parts: 

\noindent {\bf Part I: Introduction to Logic}
\begin{itemize}

\item {\em Informal laws of mathematical reasoning}

Our starting point is the place where the students left off in a discrete mathematics course: with basic set-theoretical concepts. However, our primary focus is not on understanding the concepts themselves, but on {\em reasoning} about them by applying informal logical laws.  Accordingly, the students are asked to provide proofs of basic claims, explaining which laws were used at each stage. The presentation of the informal laws and other proof tips is adapted from \cite{makinson2012sets}. The informal laws become explicit at the object level when classical propositional and first-order logic are introduced to the students (E.g., the law for proving general statements can be captured by the rule inferring $\forall x \psi$ from $\psi(x)$, and the law for proving conditional statements is captured by the deduction theorem.) %At this stage we revisit the proofs and pinpoint the application of these laws. 

\item {\em Induction:} mathematical, structural and computational induction.  \\
Induction is in the heart of several formal concepts relevant for verification and validation of software: fixed point constructions, model checking, program analysis and many more. Therefore a special emphasis is put on the topic throughout the course, highlighting its various manifestations, e.g. proving syntactic properties of logical formulas, proving the deduction theorem, proving invariants with respect to a Z specification. Software-related examples are adapted from Chapter 2 of  (\cite{aho1992foundations}).

\item {\em Classical Propositional and First-Order Logic:} syntax and semantics, satisfiability and validity, Hilbert-style axiomatization, formalization of natural language sentences. \\
For this part of the course we mostly adapt parts of the standard presentation of most mathematical logic textbooks. We make a special emphasis on formalization of natural language specifications and induction at the expense of omitting the proofs of the Completeness and Compactness theorems (in line with the recommendation of \cite{makowsky2008hilbert}).

\item {\em Survey of non-classical logics\footnote{This part of the course is implemented by assigning each of the students a short presentation on a non-classical logic or its applications of his choice. While the importance of temporal logic in this context is perhaps the most obvious one due to its well-known applications in verification, also other non-classical logics have IS-relevant applications. Our goal here is to increase the awareness of the students to the immense variety of logics outside the realm of classical logics, as well as engage them more actively in the course. Several students have reported that exploring new logics on their own was the part they enjoyed the most in the course.       
}:} temporal logic, modal logic, many-valued logic, fuzzy logic, non-monotonic logic, paraconsistent logic.

%\item {\em First-order logic (FOL):} syntax and semantics, quantification, Hilbert-style axiomatization, satisfiability and validity, formalization of software requirements. 

\end{itemize}

\noindent {\bf Part II: Introduction to Formal Specification}

This part of the course builds up on the knowledge obtained at the previous part. The final aim is for the students to be able to understand and write formal specifications using the Z notation. For this we have adapted the material from the textbook \cite{potter1996introduction}, covering the basic aspects of Z: types, schemas and reasoning about Z specifications. However, staying faithful to our vision oulined above, we have developed our own set of examples, which are (i) "simple" and (ii) related to software domains. In what follows we shortly discuss what we mean by "simple" and how "simplicity" can be measured. 

\subsubsection*{Measuring comprehensibility of Z specifications}
The notion of simplicity (item (i) above) is not well understood. Comprehensibility (or understandability) of specifications is usually thought of as the degree to which information contained in a specification is understandable to the reader, and this is a well-studied topic in software engineering (see, e.g., \cite{condori2011practical} for a survey). However, we are aware of only a few works on Z specifications (\cite{finney1999effects,finney1998measuring}), all of which on the structural dimension. We believe, however, that simplicity is a key to comprehensibility of specifications, at least at the stage of learning the topic. Our attempt to quantitatively measure "simplicity has led to our ongoing project of developing automatic tools for this purpose. Our current hypothesis is that the {\em nesting} of definitions is and shortening notations by {\em introducing additional symbols} decrease the understandability of specifications. We are currently developing a tool\footnote{The tool is based on the open-source Java framework Community Z tools (\cite{Ztools}).} for measuring "simplicity" of specifications. Using this tool, we plan to empirically check our hypothesis, as well as to consider other comprehensibility dimensions.

\subsubsection*{Students' Acceptance}

The course has only been taught in its current form for five years, so making decisive conclusions about its effectiveness is perhaps premature. However, an important dimension in evaluating such effectiveness is the students' acceptance and reaction. To gain a better understanding of these factors, the first author has undertaken a preliminary qualitative study using a questionnaire filled by twenty three students who took the course in 2013-2014.  

It should be noted that the limiting factors typical of our target audience are in many aspects similar to those described in \cite{tavolato2012integrating}. The first is {\em lack of mathematical background}: the undergraduate IS study program at the University of Haifa does not include a course in logic, and the majority of students have only a background in discrete mathematics, where they are taught very basic concepts of set theory. The second limiting factor is their {\em lack of motivation}: the majority of the students return to graduate school several years after receiving their B.A, while working full-time. They typically expect the topics to be directly relevant to their IS practice, and usually exhibit difficulty in coping with the dense and abstract material taught in the course. In light of these factors, we were expecting some of the students to claim, basically, that the course was too hard without being helpful for their future as IS practitioners. However, only one student out of 23 felt the course was not useful for his practice.   
A full analysis of the data obtained from students is out of the scope of this paper. However, some aspects highlighted by the data were that (i) the majority of students felt the course has improved their analytical thinking abilities\footnote{Examples: {\sf "I don't use logic or Z on a daily basis in my research or my work, but it improved my modelling skills", ``I think it can be even more useful as an introduction to programming, because it teaches you to think systematically", {\sf "Every graduate student needs this course as a basis for study and research"}.}}; (ii) the majority of students felt the part related to formal specification is relevant for their IS practice\footnote{Examples: {\sf ``I think Z language is the most useful part, because it is applied in industry"}, {\sf ``If I encounter any other formal notation in industry, it will take me less time to get familiar with the subject"}.}; and (iii) some students felt the course has {\em directly} improved their daily IS practice\footnote{Examples: {\sf "I started using truth tables at work to rule out impossible behaviors", "I was surprised to find out how helpful the tools we obtained are in my daily work".}}.

\section{Summary and Future Research}

There has recently been a discourse on the relevance of traditional logic courses to future computer science practitioners. Jeannette Wing writes in  \cite{DBLP:conf/amast/Wing95}: ``...we still face the educational challenge of teaching mathematical foundations like logic and discrete mathematics to practicing or aspiring software engineers. {\em We need to go beyond giving the traditional courses and think about who the target students are.}" This paper discusses these issues for the target population of IS students, for whom the lack of direct relevance of the traditional logic courses seems to have led to their exclusion from the curriculum. We believe logic is central to IS objectives, as it is the key to applying formal methods in specification, verification and validation of information systems. Ideally, we need more empirical evidence in the spirit of the Beseme project (\cite{Beseme}) that such courses are useful for IS practitioners. In addidion, there is a need for a wider discussion on {\em what} logical background is needed for Information Systems practitioners and {\em how} it should be taught. In a contribution to such discussion, we have reported on our insights from teaching the ``Logic and Formal Specification" course to graduate IS students. Like previous authors report in the context of CS, we have seen that using software-related and comprehensible examples, as well as simplification of logical intricacies contributes to achieving the courses' objectives. From a more practical perspective, a future direction is an empirical investigation of {\em how} to make formal specification more understandable for students. This question is particularly interesting due to its direct relation to the more general topic of comprehensibility of specifications. In this context we plan to extend and refine our tool for automatic analysis of Z specifications and carry out an empirical evaluation. From the angle of education, strategies for an efficient integration of logic and formal methods into the IS curricula are required (along the lines of \cite{barland2000integrating,wing2000invited}), as well as an investigation of the ways to bridge intuitive and analytical thinking processes in logic and formal methods (along the lines of \cite{ejersbo2014bridging}). In this context we would also like to point out that a textbook with an IS-orientation would be a welcome addition to the large existing variety of CS-oriented books.

\bibliography{teaching}

\end{document}